# Acceleration Strategies for MR-STAT: Achieving High-Resolution Reconstructions on a Desktop PC within 3 minutes

Hongyan Liu, Oscar van der Heide, Stefano Mandija, Cornelis A. T. van den Berg, Alessandro Sbrizzi

*Abstract*— MR-STAT is an emerging quantitative magnetic resonance imaging technique which aims at obtaining multi-parametric tissue parameter maps from single short scans. It describes the relationship between the spatial-domain tissue parameters and the time-domain measured signal by using a comprehensive, volumetric forward model. The MR-STAT reconstruction solves a large-scale nonlinear problem, thus is very computationally challenging. In previous work, MR-STAT reconstruction using Cartesian readout data was accelerated by approximating the Hessian matrix with sparse, banded blocks, and can be done on high performance CPU clusters with tens of minutes. In the current work, we propose an accelerated Cartesian MR-STAT algorithm incorporating two different strategies: firstly, a neural network is trained as a fast surrogate to learn the magnetization signal not only in the full time-domain but also in the compressed low-rank domain; secondly, based on the surrogate model, the Cartesian MR-STAT problem is re-formulated and split into smaller sub-problems by the alternating direction method of multipliers. The proposed method substantially reduces the computational requirements for runtime and memory. Simulated and in-vivo balanced MR-STAT experiments show similar reconstruction results using the proposed algorithm compared to the previous sparse Hessian method, and the reconstruction times are at least 40 times shorter. Incorporating sensitivity encoding and regularization terms is straightforward, and allows for better image quality with a negligible increase in reconstruction time. The proposed algorithm could reconstruct both balanced and gradient-spoiled in-vivo data within 3 minutes on a desktop PC, and could thereby facilitate the translation of MR-STAT in clinical settings.

## I. INTRODUCTION

MAGNETIC Resonance Spin Tomography in Time-Domain ("MR-STAT") [1] is a recently developed framework using model-based iterative reconstruction techniques for simultaneously obtaining multiple parametric quantitative tissue maps, such as $T_1, T_2$ and proton density ($\rho$), from a single short scan. One main feature of the MR-STAT reconstruction is the application of a Bloch-equation based volumetric signal model and the direct fitting of the spatial parameter maps to the measured temporal-domain signal. Most quantitative MR reconstructions, for example the MR Fingerprinting (MRF) technique [2], use a two-step approach to reconstruct tissue maps consisting of (a) a spatial localization step based on Fourier transform, and (b) a voxel-wise parameter quantification step based on the physical model. Unlike those two-step approaches, MR-STAT describes the comprehensive relationship between the measured time-domain signal and the spatial tissue parameter maps as a large-scale nonlinear problem [3]–[5]; the parameter maps are reconstructed by derivative-based iterative optimization algorithms. With this approach, data from different k-space readouts are combined and processed in one single reconstruction step; therefore, MR-STAT reconstructions do not suffer from FFT-induced aliasing artifacts even when using a very short transient-state sequence with standard Cartesian acquisition strategies [6].

Although MR-STAT has been demonstrated for reconstructing high-quality parameter maps, the MR-STAT reconstruction is a computationally demanding problem for the following two reasons. Firstly, as a derivative-based iterative method, the MR-STAT reconstruction requires the computation of MR signals for each voxel of the image at each iteration as well as their derivatives with respect to the tissue parameters, thus leading to long computation time [6]. Secondly, since the MR-STAT reconstruction solves a large-scale nonlinear problem in which many unknown parameters (usually in the order of $10^5$ ) are updated simultaneously, memory requirements can be very high.

Recent efforts have been made to tackle the computationally challenging MR-STAT reconstruction problem. In a recent work [7], a highly parallelized Gauss-Newton reconstruction algorithm using the sparse Hessian approximation technique was presented. In the sparse Hessian algorithm, a sparse, banded structure was exploited in the Hessian matrix approximation process to reduce the memory and computational time requirements, and therefore the MR-STAT reconstruction times were substantially reduced by approximately one order of magnitude. However, reconstruction times for one single 2D slice using the sparse Hessian method are still in the order of tens of minutes, despite the use of a high performance CPU cluster with 96 cores.

Manuscript submitted to IEEE Transactions on Medical Imaging on the 22th of August 2021. This work was funded by the Chinese Scholarship Council (CSC) under grant number 201807720088.

Hongyan Liu, Oscar van der Heide, Stefano Mandija, Cornelis A. T. van den Berg, Alessandro Sbrizzi are with the Center for Image Sciences, University Medical Center, Utrecht, Heidelberglaan 100, 3584 CX The Netherlands (email: h.liu@umcutrecht.nl) .

In this work, we focus on further reduction of the reconstruction time by presenting a new algorithm which combines different strategies and can be run on a Desktop PC.

Firstly, in order to reduce the time for the computation of the MR signal and its derivatives, the Bloch equation simulations are replaced by a fast surrogate model in the form of a fully-connected neural network (NN) [8]. Inspired by the singular value decompression (SVD) based compression schemes for the temporal MR signal [9], [10], we design a network architecture which learns not only the full magnetization response and its derivatives, but also compressed low-rank signals. Currently, the application of neural network in quantitative MRI mainly focuses on the following two aspects: On one hand, various neural networks have been trained to learn the Bloch-based forward model for MR signal simulations such as MR Fingerprinting dictionary generation [11], [12]; on the other hand, deep learning methods have also been directly applied as fast inverse problem solvers for quantitative MR reconstructions [13], [14]. In this work, we use the neural network for accelerating MR-STAT reconstruction in a different way: the model-based MR-STAT reconstruction problem is still solved by an iterative nonlinear solver, whereas the forward model computation as well as the forward model derivatives are accelerated by a pre-trained neural network [15].

Secondly, assuming Cartesian gradient trajectories are used in the acquisition, the volumetric MR-STAT signal model is factorized into linear and nonlinear operators. Given the application of the surrogate model and the factorization, the resulting MR-STAT reconstruction can be split into simpler problems and then alternatively solved by the alternating direction method of multipliers (ADMM), which has previously been applied to various quantitative MR reconstruction problems [16]–[19]. In our proposed ADMM algorithm for MR-STAT, the first sub-problem is linear, and includes the phase-encoding and signal compression operators. The second sub-problem is nonlinear, but it can be separated into many smaller, independent problems. We show that the proposed algorithm is memory efficient, and drastically reduces the MR-STAT reconstruction time (40 times faster than state-of-the-art MR-STAT reconstructions), allowing for reconstructions of quantitative maps within three minutes on a standard desktop PC with 8 CPU cores.

## II. THEORY

In this section, the original MR-STAT framework is first reviewed. We then present the architecture of the neural network as a fast surrogate model for computing MR signals. Finally, we show how the surrogate model can be incorporated in the factorized MR-STAT framework to accelerate the MR-STAT reconstruction by means of ADMM schemes.

### A. MR-STAT Framework

The temporal transverse magnetization signal $m(t)$ can be computed by the solution of the Bloch equation. The magnetization evolution in time depends on both the MR-related biological tissue parameter $\alpha(r) = [T_1(r), T_2(r), \rho(r), ...]$ with two-dimensional spatial coordinates $r = [x, y]^T$, and the applied time-varying pulse sequence (e.g. radiofrequency excitation pulses and spatial encoding gradient fields) with sequence parameter denoted by $\beta$. Here $T_1(r)$ and $T_2(r)$ denote longitudinal and transverse relaxation time respectively, and $\rho(r) \in \mathbb{C}$ represents the proton density of the tissue weighted by the complex receive RF field $B_1^-$.

By applying the Faraday's law of induction [20], the volumetric time-domain signal measured from a receiver coil can be modelled as:

$$s(\alpha, t) = \int_V \rho(r) m(t; \alpha(r), \beta) dr \quad (1)$$

which is the spatial integration of the transverse magnetization of all excited spin isochromats over the whole region.

Let $d(t)$ be the signal measured from the receiver coil of the MR scanner, the MR-STAT problem can be formulated as the following optimization problem

$$\alpha^* = \arg\min_\alpha \left\{ \int_V |d(t) - s(\alpha, t)|^2 dt \right\}, \quad (2)$$

such that $s(\alpha, t)$ is modelled by equation (1) and the required $m(t; \alpha(r), \beta)$ is modelled by the Bloch equation. The optimal multi-parameter map $\alpha^*$ is obtained by numerically solving equation (2) which minimizes the difference between the physics-based computed signal $s(\alpha, t)$ and the measured signal $d(t)$.

During an MR-STAT experiment, the magnetization is no longer in steady-state, and computing the temporal signal $m(t; \alpha(r), \beta)$ for all spatial locations $r$, (usually in the order of 10^5 for 2D problems), by Bloch equation is one of the most time-consuming steps. To better understand the acceleration strategies proposed in this work, we derive a detailed MR-STAT problem formulation incorporating the signal computation based on (2). The new acceleration strategy will be explained in the following sections and will be based on this new formulation.

The full MR signal computation can be split into two steps: we first compute the magnetizations at each echo times, and then compute the MR signal evolution during each readout time period from the corresponding echo time signal.

**First-step: computation of the signal at echo times**

Let $N_{TR}$ be the number of RF excitations within a sequence, and $N_{Read}$ be the number of readout samples after each RF excitation. For the first step, let vector $m^E(\alpha(r), \beta) = [m(t_1^E; \alpha(r), \beta); m(t_2^E; \alpha(r), \beta); ....; m(t_{N_{TR}}^E; \alpha(r), \beta)] \in \mathbb{C}^{N_{TR} \times 1}$ be the complex magnetization signal at all echo times, and the echo time signal can be computed evolutionally by

$$m(t_j^E; \alpha(r), \beta) = MR\_Signal\left\{m(t_{j-1}^E; \alpha(r), \beta), \beta\right\}, \quad (3)$$
$$j = 1, 2, ..., N_{TR}$$

where $MR\_Signal\{\cdot\}$ represents the MR signal model (Extended Phase Graph, EPG, or Bloch equation), and the sequence parameter required for computing the $j$-th echo time signal includes the RF excitation pulse, the slice-selective gradient, echo time $T_E$ and repetition time $T_R$.

**Second-step: expanding the signal to all readout samples**

In the second MR signal computation step, the complete temporal signal for all samples during the readout is expanded from the echo time signal using a closed form solution of the Bloch equation. Let $t_{j,n}$ be the sample time for the $n$-th readout sample after the $j$-th RF excitation. After the $j$-th RF excitation, the gradient encoding vector $\mathbf{gr}_j(r)$, the off-resonance effect factor $\mathbf{b}_j(\alpha(r))$ and transverse signal decay factor $\mathbf{de}_j(\alpha(r))$ for spatial location $r$ are defined as follows

$$\mathbf{gr}_j(r) = \left[ e^{-ik(t_{j,1})\cdot r}, e^{-ik(t_{j,2})\cdot r}, \ldots, e^{-ik(t_{j,N_{Read}})\cdot r} \right],$$

$$\mathbf{b}_j(\alpha(r)) = \left[ e^{-i\gamma B_0(r)\cdot(t_{j,1}-t_j^E)}, e^{-i\gamma B_0(r)\cdot(t_{j,2}-t_j^E)}, \ldots, e^{-i\gamma B_0(r)\cdot(t_{j,N_{Read}}-t_j^E)} \right],$$

$$\mathbf{de}_j(\alpha(r)) = \left[ e^{-\frac{t_{j,1}-t_j^E}{T_2(r)}}, e^{-\frac{t_{j,2}-t_j^E}{T_2(r)}}, \ldots, e^{-\frac{t_{j,N_{Read}}-t_j^E}{T_2(r)}} \right],$$

(4)

while all the magnetization readouts after the $j$-th RF excitation can be computed by

$$\boldsymbol{m}_j(\alpha(r), \beta) = m(t_j^E; \alpha(r), \beta) \cdot \left[ \mathbf{gr}_j(r) \odot \mathbf{b}_j(\alpha(r)) \odot \mathbf{de}_j(\alpha(r)) \right] \in \mathbb{C}^{1 \times N_{Read}}, \quad (5)$$

where $\odot$ denotes the Hadamard product, and $k(t_{j,n}) = [k_x(t_{j,n}), k_y(t_{j,n})]$ represents the accumulating effect of the encoding gradient fields. Note that the gradient encoding vector $\mathbf{gr}_j(r)$ can be interpreted as a Fourier term by the acquisition gradients.

Let the matrix $M(\alpha(r), \beta) = [\boldsymbol{m}_1(\alpha(r), \beta), \boldsymbol{m}_2(\alpha(r), \beta), \ldots, \boldsymbol{m}_{N_{TR}}(\alpha(r), \beta)] \in \mathbb{C}^{N_{TR} \times N_{Read}}$ denote the full temporal magnetization signal for spatial location $r$. Using the two-step magnetization signal computation method (3)-(5) for computing $M(\alpha(r), \beta)$, and performing spatial discretization on (1), we obtain:

$$S(\alpha) = \sum_{a=1}^{N_x} \sum_{b=1}^{N_y} \rho(x_a, y_b) M(\alpha(x_a, y_b), \beta). \quad (6)$$

The MR-STAT problem (2) can finally be re-written in the following matrix format

$$\alpha^* = \arg\min_\alpha \left\| D - S(\alpha) \right\|_F^2, \quad (7)$$

here $\|\cdot\|_F^2$ denotes the squared Frobenius norm, $(x_a, y_b)$ are the spatial coordinates for voxel $(a, b)$, and the discretized image size is $N_x \times N_y$. Matrix $D \in \mathbb{C}^{N_{TR} \times N_{Read}}$ is a discrete matrix format of measured signal $d(t)$, and denotes all measured time-domain samples.

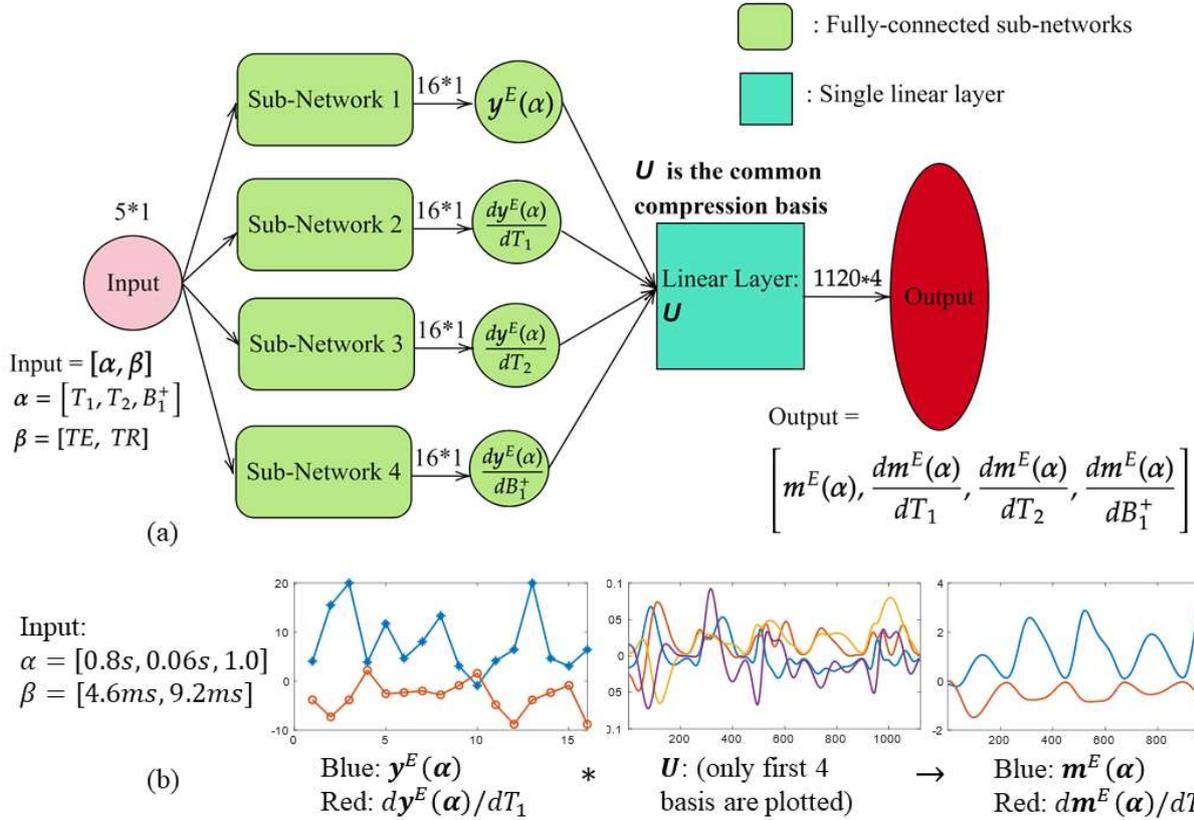

Fig. 1. (a) Neural Network (NN) architecture for the surrogate MR signal model. The input of the NN is a combination of tissue parameters ($T_1, T_2, B_1^+$ and etc.) and sequence parameters (TR, TE). The output is the time-domain signal and its derivatives w.r.t. to all tissue parameters. The NN consists of separate blocks (Sub-Networks 1 to 4) for computing the compressed signal and derivatives; each network has four fully connected layers with ReLU activation function, and one final learnable linear layer which acts as a linear decoding step by means of the learnable $U$ matrix. (b) An example of the simulated signal by (a) using the same sequence as in later Test 1 and 2.

### B. A surrogate model for MR signals and derivatives at echo time

As we mentioned in the previous section, the most intensive computations are involved in the echo-time signal calculations (equation 3). Therefore, we employ a fast surrogate model in the form of a neural network (NN), to accelerated the computation for magnetizations at echo times, that is, to approximate the function *MR_Signal*. Since MR-STAT uses a derivative-based iterative optimization scheme, both the magnetization and its derivatives need to be computed at each iteration during the reconstructions. Therefore, we design an NN which computes both the magnetization and its derivatives simultaneously.

A fully-connected NN is designed and trained to return both magnetizations and derivatives at echo times for either a balanced or gradient-spoiled sequence with time-varying flip angles. The NN architecture is shown in Fig. 1. The inputs are the spatially dependent parameters $\boldsymbol{\alpha}$, such as $T_1$, $T_2$, transmit RF field inhomogeneity $B_1^+$ and off-resonance $B_0$, and sequence parameters $\boldsymbol{\beta}$, such as $T_E$ and $T_R$. The network consists of four separate blocks (Sub-Network 1 to 4) for computing the complex compressed signals, $\boldsymbol{y}^E(\boldsymbol{\alpha}, \boldsymbol{\beta}) \in \mathbb{C}^{N_B \times 1}$, and the derivatives, such as $d\boldsymbol{y}^E/dT_1$, $d\boldsymbol{y}^E/dT_2$, $d\boldsymbol{y}^E/dB_1^+ \in \mathbb{C}^{N_B \times 1}$, where $N_B$ is the length of the compressed signal. Sub-Network 1 to 4 have the same architecture, consisting of 4 fully-connected layers, each activated by a rectified linear unit (ReLU). The outputs of Sub-Network 1 to 4 are then passed through one final linear layer to obtain the full temporal signals $\boldsymbol{m}^E(\boldsymbol{\alpha}, \boldsymbol{\beta})$, $d\boldsymbol{m}^E(\boldsymbol{\alpha}, \boldsymbol{\beta})/dT_1$, $d\boldsymbol{m}^E(\boldsymbol{\alpha}, \boldsymbol{\beta})/dT_2$ and $d\boldsymbol{m}^E(\boldsymbol{\alpha}, \boldsymbol{\beta})/dB_1^+$. Vector $\boldsymbol{m}^E \in \mathbb{C}^{N_{TR} \times 1}$ represents the complex magnetization signal at all echo times. The linear relationship between the compressed and full magnetization response can be expressed as below:

$$\boldsymbol{m}^E(\boldsymbol{\alpha}, \boldsymbol{\beta}) = \boldsymbol{U} \boldsymbol{y}^E(\boldsymbol{\alpha}, \boldsymbol{\beta}). \quad (8)$$

Here $\boldsymbol{U} \in \mathbb{C}^{N_{TR} \times N_B}$ denotes the decoding basis for the low-rank signal $\boldsymbol{y}^E(\boldsymbol{\alpha}, \boldsymbol{\beta})$, which is learned during the NN training. Existing low-rank approximations usually derive the $\boldsymbol{U}$ basis by applying SVD to a dataset of MR signals, which can be extremely memory demanding for a large dataset [10], [21]. The surrogate model proposed here learns the compression relationship during the network training, thus it avoids the memory problem. The same linear compression relationship also applies to the derivative signals. Note that the complex matrices and vectors are split into real and imaginary parts when implementing the neural network. When no $B_0$ is included and the RF flip-angle train and $B_1^+$ are real-valued, $\boldsymbol{m}^E(\boldsymbol{\alpha}, \boldsymbol{\beta})$, $\boldsymbol{y}^E(\boldsymbol{\alpha}, \boldsymbol{\beta})$ and $\boldsymbol{U}$ become real-valued quantities.

### C. A factorized MR signal model using the surrogate model

In this section, we insert the surrogate model into (6) and exploit the full factorized MR signal model for accelerating the MR-STAT reconstruction. From now on, we consider a 2D Cartesian acquisition, which is commonly used in MR-STAT. Without loss of generality, we assume the phase-encoding direction is directed along the $y$ axis and readout-encoding along the $x$ axis. For Cartesian acquisition, the gradient encoding vector $\mathbf{gr}_j(\boldsymbol{r})$ can be split into the multiplication of two terms, a scalar phase-encoding factor $\mathrm{ph}_j(\boldsymbol{r})$ which is independent of the readout index $n$, and a frequency encoding vector $\mathbf{fr}(\boldsymbol{r})$ which is independent of the RF index $j$, defined as below

$$\begin{aligned}
\mathrm{ph}_j(\boldsymbol{r}) &= e^{-ik_y(t_j^E) \cdot y}, \\
\mathbf{fr}(\boldsymbol{r}) &= \left[ e^{-ik_x(\Delta t_1) \cdot x}, e^{-ik_x(\Delta t_2) \cdot x}, \ldots e^{-ik_x(\Delta t_{N_{Read}}) \cdot x} \right].
\end{aligned} \quad (9)$$

Here, $\Delta t_n = t_{j,n} - t_j^E$ is the time between the time for the $n$-th readout sample and the echo time, and the time interval $\Delta t_n$ is typically independent of RF index $j$ for Cartesian sequences because readout sampling times are the same after each RF excitations. Equation (5) can be rewritten by using (9) and re-ordering the terms

$$\begin{aligned}
\boldsymbol{m}_j(\boldsymbol{\alpha}(\boldsymbol{r}), \boldsymbol{\beta}) &= \mathrm{ph}_j(\boldsymbol{r}) \cdot m(t_j^E; \boldsymbol{\alpha}(\boldsymbol{r}), \boldsymbol{\beta}) \cdot \\
&\quad \left[ \mathbf{b}_j(\boldsymbol{\alpha}(\boldsymbol{r})) \odot \mathbf{de}_j(\boldsymbol{\alpha}(\boldsymbol{r})) \odot \mathbf{fr}(\boldsymbol{r}) \right] \in \mathbb{C}^{1 \times N_{Read}}.
\end{aligned} \quad (10)$$

Equation (10) computes the magnetization at all readout times after the $j$-th RF excitation. The full magnetization after all RF excitations can be computed by (10) for $j \in [1, \ldots N_{TR}]$ and concatenated in a matrix format as

$$\boldsymbol{M}(\boldsymbol{\alpha}(\boldsymbol{r}), \boldsymbol{\beta}) = \boldsymbol{C}^P(y) \cdot \boldsymbol{m}^E(\boldsymbol{\alpha}(\boldsymbol{r}), \boldsymbol{\beta}) \cdot \boldsymbol{c}^R(\boldsymbol{\alpha}(\boldsymbol{r})), \quad (11)$$

where the matrices $\boldsymbol{C}^P$ and $\boldsymbol{M}$ and the vectors $\boldsymbol{m}^E$ and $\boldsymbol{c}^R$ are defined in Table I.

By inserting equation (11) into the volumetric signal equation (6), a new volumetric signal model is derived which makes use of the surrogate model $\boldsymbol{m}^E(\boldsymbol{\alpha}, \boldsymbol{\beta})$:

$$\begin{aligned}
\boldsymbol{S}(\boldsymbol{\alpha}) &= \sum_{a=1}^{N_x} \sum_{b=1}^{N_y} \rho_{a,b} \cdot \boldsymbol{C}^P(y_b) \cdot \boldsymbol{m}^E(\boldsymbol{\alpha}_{a,b}, \boldsymbol{\beta}) \cdot \boldsymbol{c}^R(\boldsymbol{\alpha}_{a,b}) \\
&= \sum_{b=1}^{N_y} \boldsymbol{C}^P(y_b) \cdot \\
&\quad \left[ \boldsymbol{m}^E(\boldsymbol{\alpha}_{1,b}, \boldsymbol{\beta}) \quad \boldsymbol{m}^E(\boldsymbol{\alpha}_{2,b}, \boldsymbol{\beta}) \quad \ldots \quad \boldsymbol{m}^E(\boldsymbol{\alpha}_{N_x,b}, \boldsymbol{\beta}) \right] \cdot \\
&\quad \begin{bmatrix} \rho_{1,b} \cdot \boldsymbol{c}^R(\boldsymbol{\alpha}_{1,b}) \\ \rho_{2,b} \cdot \boldsymbol{c}^R(\boldsymbol{\alpha}_{2,b}) \\ \vdots \\ \rho_{N_x,b} \cdot \boldsymbol{c}^R(\boldsymbol{\alpha}_{N_x,b}) \end{bmatrix} \\
&= \sum_{b=1}^{N_y} \boldsymbol{C}^P(y_b) \cdot \boldsymbol{M}^E(\boldsymbol{\alpha}_b^R) \cdot \boldsymbol{C}^R(\boldsymbol{\alpha}_b^R) \\
&= \sum_{b=1}^{N_y} \boldsymbol{C}^P(y_b) \cdot \boldsymbol{U} \cdot \boldsymbol{Y}^E(\boldsymbol{\alpha}_b^R) \cdot \boldsymbol{C}^R(\boldsymbol{\alpha}_b^R).
\end{aligned} \quad (12)$$

Here, $\boldsymbol{\alpha}_{a,b}$ denotes the spatial parameters for the discretized voxel at $(a, b)$, and $\boldsymbol{\alpha}_b^R = [\boldsymbol{\alpha}_{1,b}, \boldsymbol{\alpha}_{2,b}, \ldots, \boldsymbol{\alpha}_{N_x,b}]$ denotes all spatial parameters for one row of voxels along the readout direction and $\rho(r) \in \mathbb{C}$ denotes the proton density of the tissue weighted by the complex receive RF field $B_1^-$. From line 2 to line 5, summation over $a$ is computed by concatenating $m^E(\alpha_{a,b},\beta)$ and $c^R(\alpha_{a,b})$ vectors into matrices $M^E(\alpha_b^R)$ and $C^R(\alpha_b^R)$, where $M^E(\alpha_b^R) \in \mathbb{C}^{N_{TR} \times N_x}$ includes all the signals at echo times for the $b$-th phase encoding voxel line, and $C^R(\alpha_b^R) \in \mathbb{C}^{N_x \times N_{Read}}$ is the readout effect matrix. By applying the linear relationship (8), we obtain the final line in (12). Here, $Y^E(\alpha_b^R) \in \mathbb{C}^{N_B \times N_x}$ includes all the compressed signals at echo times for the $b$-th phase encoding voxel line, and it can be computed by the surrogate model (Sub-Network 1 in Fig. 1) described in the previous sub-section.

TABLE I
EXPLANATION OF MATRICES AND VECTORS USED IN (12).

| Matrix/vector and definition | Size | Description |
|---|---|---|
| $M(\alpha(r),\beta) = \begin{pmatrix} m(t_{1,1};\alpha(r),\beta) & \cdots & m(t_{1,N_{Read}};\alpha(r),\beta) \\ \vdots & \ddots & \vdots \\ m(t_{N_{TR},1};\alpha(r),\beta) & \cdots & m(t_{N_{TR},N_{Read}};\alpha(r),\beta) \end{pmatrix}$ | $\mathbb{C}^{N_{TR} \times N_{Read}}$ | A matrix including all readout signals for one fixed voxel |
| $C^P(y) = diag\{(\text{ph}_1(r),\ldots,\text{ph}_{N_{TR}}(r))\}$ | $\mathbb{C}^{N_{TR} \times N_{TR}}$ | A diagonal matrix describing the phase-encoding effect |
| $m^E(\alpha(r),\beta) = (m_1^E(\alpha(r),\beta),\ldots,m_{N_{TR}}^E(\alpha(r),\beta))^T$ | $\mathbb{C}^{N_{TR} \times 1}$ | A column vector representing magnetization signals at all echo times |
| $c^R(\alpha(r)) = [\mathbf{b}_j(\alpha(r)) \odot \mathbf{de}_j(\alpha(r)) \odot \mathbf{fr}(r)]$ | $\mathbb{C}^{1 \times N_{Read}}$ | A row vector including all readout effects such as frequency encoding, transverse decay and off-resonance |

The final result in (12) shows that the volumetric temporal signal $S(\alpha)$ can be computed by adding all MR signals from each row of voxels in the phase-encoding ($y$) direction; the MR signal from one row of the voxels is computed by the multiplication of four different matrix operators. A graphic illustration of this new signal model and the explanation of the operators are shown in Fig. 2.

### D. Alternating Directions Method of Multipliers (ADMM)

Inserting the new factorized MR signal model (12) into (7) leads to the following new format of the MR-STAT problem:

$$\alpha^* = \arg\min_\alpha \frac{1}{2} \left\| D - \sum_{b=1}^{N_y} C^P(y_b) \cdot U \cdot Y^E(\alpha_b^R) \cdot C^R(\alpha_b^R) \right\|_F^2. \quad (13)$$

Noticing that in (13), the nonlinear dependency on $\alpha$ only applies to the matrices $Y^E(\alpha_b^R)$ and $C^R(\alpha_b^R)$, we reformulate problem (13) as a linear optimization problem with nonlinear constraints,

$$\alpha^* = \arg\min_\alpha \frac{1}{2} \left\| D - \sum_{b=1}^{N_y} C^P(y_b) \cdot U \cdot Z_b \right\|_F^2$$

$$\text{subject to } Y^E(\alpha_b^R) \cdot C^R(\alpha_b^R) - Z_b = 0 \quad (14)$$

$$\text{for } b \in [1,\ldots,N_y],$$

where the auxiliary variable $Z_b \in \mathbb{C}^{N_B \times N_{Read}}$ represents the compressed signal for the $b$-th row of voxels excluding the phase encoding gradient effects. Notice that (14) is a linear optimization problem in $Z$ with $N_y$ nonlinear independent constraints.

We apply ADMM to the MR-STAT reconstruction problem (14) via variable splitting. This approach can be realized by an augmented Lagrangian in a scaled form [16]:

$$\{\alpha_{ADMM}^*, Z_{ADMM}^*, W_{ADMM}^*\} = \arg\min_{\alpha,Z,W} \mathcal{L}_\lambda(\alpha,Z,W)$$

$$= \arg\min_{\alpha,Z,W} \left\{ \frac{1}{2} \left\| D - \sum_{b=1}^{N_y} C^P(y_b) \cdot U \cdot Z_b \right\|_F^2 \right. \quad (15)$$

$$+ \frac{\lambda}{2} \sum_{b=1}^{N_y} \left\| Y^E(\alpha_b^R) \cdot C^R(\alpha_b^R) - Z_b + W_b \right\|_F^2$$

$$\left. - \frac{\lambda}{2} \sum_{b=1}^{N_y} \|W_b\|_F^2 \right\}.$$

Here, the first term models the data consistency, and the second term models the nonlinear properties of the MR signal. $W_b$ for $b \in [1, N_y]$ is the scaled dual variable and $\lambda$ is the ADMM penalty parameter. The ADMM algorithm for solving (15) is shown in Algorithm 1, which alternatively updates $\alpha, Z, W$ in each iteration.

**Algorithm 1.** Accelerated MR-STAT

Solve problem (15) by an ADMM algorithm.
**Require**: Initial guess $\alpha^{(0)}$, $Z^{(0)}$ and $W^{(0)}$.
for j=1,2,3,….
1. Linear step solution:

$$Z^{(j+1)} = \arg\min_{Z} \mathcal{L}_\lambda\left(\alpha^{(j)}, Z, W^{(j)}\right)$$

$$= \arg\min_{Z} \left\{ \frac{1}{2} \left\| D - \sum_{b=1}^{N_y} C^P(y_b) \cdot U \cdot Z_b \right\|_F^2 \right. \quad (16)$$

$$\left. + \frac{\lambda}{2} \sum_{b=1}^{N_y} \left\| Y^E\left(\alpha_b^{R,(j)}\right) \cdot C^R\left(\alpha_b^{R,(j)}\right) - Z_b + W_b^{(j)} \right\|_F^2 \right\},$$

2. Nonlinear step solution:

$$\alpha^{(j+1)} = \arg\min_{\alpha} \mathcal{L}_\lambda\left(\alpha, Z^{(j+1)}, W^{(j)}\right)$$

$$= \arg\min_{\alpha_b^R} \left\| Y\left(\alpha_b^R\right) C^R\left(\alpha_b^R\right) - Z_b^{(j+1)} + W_b^{(j)} \right\|_F^2 \quad (17)$$

$$\text{for } b \in \left[1, N_y\right],$$

3. Lagrangian multiplier update:

$$W_b^{(j+1)} = W_b^{(j)} +$$
$$Y^E\left(\alpha_b^{R,(j+1)}\right) \cdot C^R\left(\alpha_b^{R,(j+1)}\right) - Z_b^{(j+1)}. \quad (18)$$

**end for**

In the accelerated Algorithm 1, Equation (16) is a linear optimization problem and a closed form solution can be computed. In (17), the nonlinear problem can be easily split into $N_y$ separate nonlinear sub-problems, and they can be solved by a trust-region method. Equation (18) simply updates the Lagrangian multiplier used for ADMM algorithms. In order to better ensure the convergence of the MR-STAT problem, we initialize the algorithm with constant tissue parameters $\alpha^{(0)}$ and with an FFT-based image mask, $Z^{(0)} = 0$ and $W^{(0)} = 0$. The image mask is acquired by applying a magnitude threshold to the FFT-reconstructed contrast image of the acquired k-space data. In MR-STAT, subsequently acquired read-out lines eventually fill up a full k-space, thus FFT could be applied to obtain anatomical information. During the iterations to solve (17), compressed signals $Y(\alpha_b^R)$ and the corresponding derivatives $\partial Y(\alpha_b^R)/\partial \alpha_b^R$ are quickly computed by the surrogate model developed in the previous sub-section.

Note that the nonlinear sub-problem (17) consists of $N_y$ independent small problems, each of which updates tissue parameters for one line of voxels in the readout-encoding direction. These small independent problems can be solved in parallel, leading to additional acceleration. The total computation time for accelerated MR-STAT is approximately in the order of $\mathcal{O}(N_y \cdot N_x^2)$, because the most time-consuming step is the solution of the $N_y$ nonlinear subproblems (equation 17), and the computational time for each nonlinear subproblem is in the order of $\mathcal{O}(N_x^2)$.

The implementation of ADMM splitting in Algorithm 1 also reduces the memory requirements for MR-STAT reconstructions. When solving the large-scale MR-STAT problem (7) or (13) without this splitting strategy, derivative-based optimization algorithm requires the computation of a Jacobian matrix with size of $(N_{TR} \cdot N_{Read}) \times (N_x \cdot N_y)$, which is usually in the order of $10^6 \times 10^5$. However, when combining the surrogate model and the ADMM splitting, each row of voxels are updated independently in the nonlinear step update (18) using the compressed signal, and the size of Jacobian matrices required for solving these small nonlinear problems can be reduced to $(N_B \cdot N_{Read}) \times N_y$, usually in the order of $10^4 \times 10^3$, thus 4 order of magnitude smaller that the full

$C^P(y_b)$ : Phase-encoding operator;
$U$ : Compression basis from NN training;
$Y^E(\alpha_b^R)$: Compressed signal from the surrogate model ;
$C^R(\alpha_b^R)$: Readout effect operator (Frequency-encoding, $T_2$ decay, off-resonance and etc);

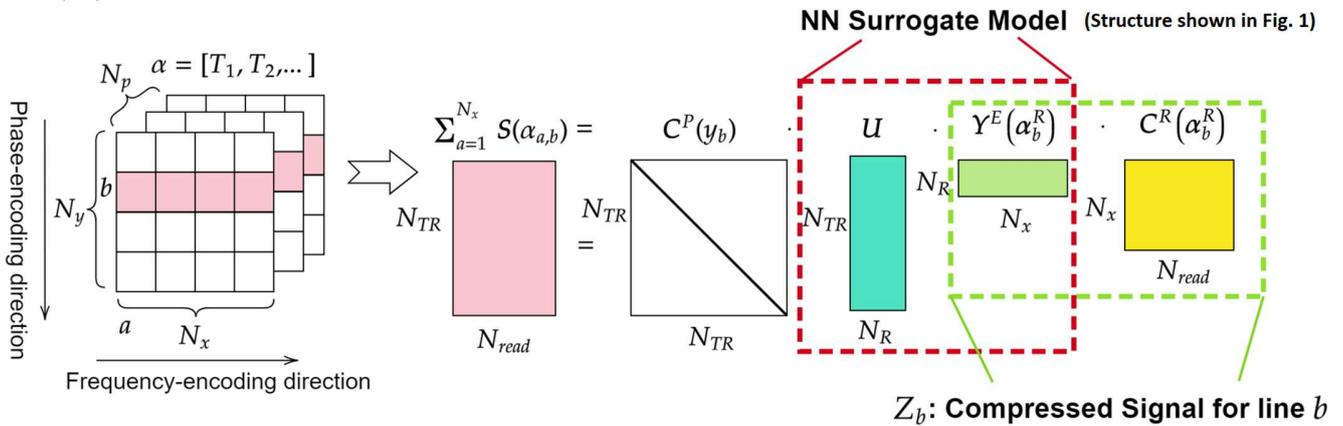

Fig. 2. Graphic illustration of the full MR signal model (12) using the surrogate model defined in Fig. 1. The graph shows the full MR signal from one frequency-encoding line (pink in the left) can be factorized into four operators, $C^P(y_b)$, $U$, $Y^E(\alpha_b^R)$ and $C^R(\alpha_b^R)$, with definitions given in the figure and in the text. $N_p$ is the number of fully sampled k-spaces. Algorithm 1 shows how problem (13) incorporating this factorized signal model can be solved using an ADMM scheme.

Jacobian. The sparse Hessian approximation can also reduce the memory requirements of the Hessian matrix but the reduction rate depends on the applied sequence. [7]

### 1) Regularization

In order to achieve better image quality for the reconstructed parameter maps, a regularization term, $R(\boldsymbol{\alpha})$, can be easily added to problem (15):

$$\{\boldsymbol{\alpha}^*_{Regu}, \boldsymbol{Z}^*_{Regu}, \boldsymbol{W}^*_{Regu}\} = \arg\min_{\boldsymbol{\alpha}, \boldsymbol{Z}, \boldsymbol{W}} \{\mathcal{L}_\lambda(\boldsymbol{\alpha}, \boldsymbol{Z}, \boldsymbol{W}) + \eta R(\boldsymbol{\alpha})\}, \quad (19)$$

where $\eta$ is the regularization weight, and $R(\boldsymbol{\alpha})$ is the regularization functional, for example, it can be a Total Variation (TV) regularization term for the $T_2$ map, or TV regularization terms for all parameter maps.

In order to make corresponding nonlinear sub-steps still separable, problem (13) is reformulated as follows by adding another auxiliary variable $\boldsymbol{\theta}$, and applying the augmented Lagrangian approach again,

$$\{\boldsymbol{\alpha}^*_{Regu}, \boldsymbol{Z}^*_{Regu}, \boldsymbol{W}^*_{Regu}, \boldsymbol{\theta}^*_{Regu}\}$$
$$= \arg\min_{\boldsymbol{\alpha}, \boldsymbol{Z}, \boldsymbol{W}, \boldsymbol{\theta}} \{\mathcal{L}_\lambda(\boldsymbol{\alpha}, \boldsymbol{Z}, \boldsymbol{W}) + \eta R(\boldsymbol{\theta})\}$$
$$\text{subject to } \boldsymbol{\theta} - \boldsymbol{\alpha} = \boldsymbol{0} \quad (20)$$
$$= \arg\min_{\boldsymbol{\alpha}, \boldsymbol{Z}, \boldsymbol{W}, \boldsymbol{\theta}, \boldsymbol{V}} \left\{ \mathcal{L}_\lambda(\boldsymbol{\alpha}, \boldsymbol{Z}, \boldsymbol{W}) + \eta R(\boldsymbol{\theta}) + \frac{\varepsilon}{2} \left( \|\boldsymbol{\theta} - \boldsymbol{\alpha} + \boldsymbol{V}\|_F^2 - \|\boldsymbol{V}\|_F^2 \right) \right\}.$$

The ADMM algorithm can still be used for solving (20), and $\boldsymbol{Z}, \boldsymbol{\theta}, \boldsymbol{\alpha}, \boldsymbol{V}$ and $\boldsymbol{W}$ are alternately updated in one iteration. The additional regularization terms have negligible impact on the computation time (see the Results section).

## III. METHODS

The proposed algorithm is validated on simulated data and in-vivo acquired brain data of two healthy volunteers. Informed consent was obtained from both subjects before the MRI scan.

### A. Acquisition Scheme

For all the test cases, two 2D MR-STAT sequences, either balanced or gradient-spoiled, were used [6], [22] . Both MR-STAT sequences are transient-state sequences with the same smoothly varying flip-angle train, Linear, Cartesian sampling was applied and 1120 readout lines were acquired, which fill up five full k-spaces. These were acquired with constant TEs and TRs: for balanced sequence TE / TR = 4.6/9.2 ms and for gradient-spoiled sequence, TE / TR = 4.9/8.7 ms. The field-of-view was set to 224mm × 224mm with a spatial resolution of 1mm × 1mm × 3mm. A non-selective inversion pulse was applied at the beginning, and the total sequence duration was 10.3s for balanced sequence and 9.8s for gradient-spoiled sequence.

The balanced MR-STAT sequence was used for both the numerical simulation experiment (Test 1, explained as below) and the in-vivo experiment (Test 2), whereas the gradient-spoiled sequence was used an additional in-vivo experiment (Test 3). Balanced and gradient-spoiled in-vivo brain data were acquired on a 1.5 Tesla and 3.0 Tesla clinical MR system respectively (Ingenia, Philips Healthcare, Best, The Netherlands).

### B. Surrogate Model

In order to run the proposed algorithm, surrogate models as described in the Theory Section (Fig. 1) need to be trained in advance. Two surrogate models using the same network architecture were trained for the balanced and gradient-spoiled sequence, respectively. Network parameters were chosen to balance robustness and expressivity. Specifically, we set the length of the compressed signal, which is the dimension of the latent variable $y$, to $N_B = 16$. Sub-Networks 1-4 (see Fig. 1) have 4 fully-connected layers, with 64, 64, 64 and 16 units, respectively.

To train the surrogate model for the balanced MR-STAT sequence, a training data set with 20,000 signals was computed by Bloch equation simulations including the slice profile responses. Each signal consisted of 1120 MR magnetizations at all echo times, with T1 and T2 values randomly sampled from a uniform logarithmic distributions ranging within [0.1, 5]s and [0.01, 2]s respectively. Normalized B1 values ranged within [0.8,1.2] a.u. A validation data set with 1500 signals was also generated following the same input parameter distribution. The surrogate model was built and trained using Tensorflow 2.2 [23] on a Tesla V100 GPU with an Intel Xeon 2.6GHz processor. The training was run for 4000 epochs by an ADAM optimizer with adaptive learning rates [24], with a batch size of 200. An L2 loss function, was used during the training, and both NRMSE of the signal and its derivatives were weighted equally in the loss function. The surrogate model for the gradient-spoiled sequence was trained similarly as the balanced one, using the Extended Phase Graph (EPG) model [25] to generate datasets for training.

### C. Algorithm Implementation

The proposed ADMM Algorithm 1 was implemented in MATLAB R2016b (The MathWorks, Natick, MA, USA). The trained surrogate model (see previous paragraph) was run from Matlab. For all experiments, the reconstructed parameters were $T_1, T_2, Re(\rho)$, and $Im(\rho)$, with an assumption of homogeneous $B_1^+ = 1.0$ a.u. , and $B_0 = 0$ Hz. The proposed algorithm was compared with the state-of-the-art sparse Hessian MR-STAT reconstruction as reported in [7]. All reconstructions using both algorithms were run on an 8-core desktop PC with 3.7GHz CPUs.

### D. Experiments

#### 1) Test 1: Numerical simulation experiment

The balanced MR-STAT sequence was used for the numerical experiment. A numerical brain phantom [26] of size 224 × 224 was chosen to create simulated data sets with typical tissue parameter values. The "ground true" values of $T_1, T_2$ and $\rho$ for the main three tissue types are: for white matter, $T_1^{org} = 0.5s$, $T_2^{org} = 0.07s$, $\rho^{org} = 0.77$; for gray matter, $T_1^{org} = 0.833s$, $T_2^{org} = 0.083s$, $\rho^{org} = 0.86$; and for cerebrospinal fluid, $T_1^{org} = 2.569s$, $T_2^{org} = 0.329s$, $\rho^{org} = 1.0$. The Bloch equation model [1] was used to simulate the measured volumetric temporal signal (see (4) in Theory Section). Complex Gaussian noise was added to the simulated signal such that $SNR = \|\text{signal}\|_2/\|\text{noise}\|_2 = 50$. Fifteen iterations of the proposed ADMM algorithm were run for reconstructing the synthetic data, and different ADMM parameters $\lambda$ =0.25, 0.5, 1.0, 2.0, 4.0 were tested. The values of the original MR-STAT objective function (6) without applying the surrogate model were also computed for all iterations, in order to compare with the sparse-Hessian reconstruction results.

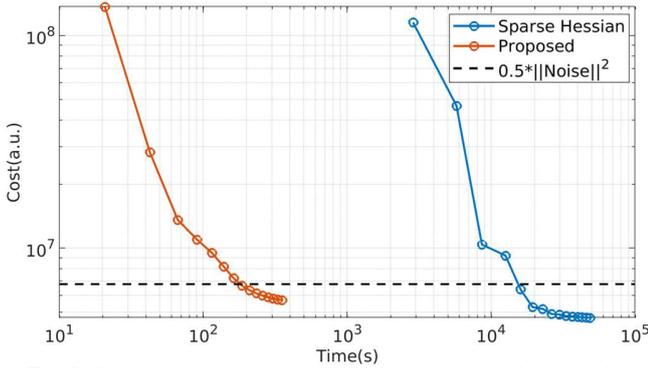

Fig. 3. Convergence curves for the numerical simulation experiment in terms of objective function versus runtime are shown. Each circle corresponds to an iteration. Note the two orders of magnitude acceleration of the proposed method when compared to the state-of-the-art sparse Hessian method. The black dashed horizontal line indicates the thermal noise level.

### 2) Test 2: In-vivo brain data with the balanced sequence

The same balanced sequence as in the previous numerical simulation experiment was used for acquiring the 2D in-vivo brain data using the manufacturer's 13 channel receiver head coil. This multi-coil data was reduced to a single virtual coil channel data via SVD compression [27]. Reconstruction was run by both the proposed ADMM algorithm and the sparse Hessian MR-STAT algorithm for comparison. Empirically, we set the ADMM parameter $\lambda$ =4.0. The same surrogate model and algorithmic settings as in the numerical simulation experiment were used for the proposed ADMM reconstruction.

### 3) Test 3: In-vivo brain data with the gradient-spoiled sequence

In order to show that the proposed algorithm could work on different types of MR-STAT sequences, we also acquired in-vivo brain data with the gradient-spoiled MR-STAT sequence. To achieve better SNR of the reconstructed images, data acquired from the 13-channel receiver head coil was reduced to 4 virtual coil channels via SVD compression, and the sensitivity profiles of the 4 virtual coils (See $g^{[j]}(r), j = 1,2,3,4$ in **Appendix A**) were also employed for reconstruction [27]. Empirically, we choose the ADMM parameter $\lambda$ =2.0. Total Variation (TV) regularizations for $T_2, Re(\rho)$, and $Im(\rho)$ were added in the reconstruction to further improve the image quality (see (19)), and the overall regularization weight $\eta$ and the corresponding ADMM parameter $\varepsilon$ are empirically chosen to be 15000 and 30000. Given the good quality of regularization-free reconstructed $T_1$ maps, we chose to add regularizations only to $T_2$ and proton density, and weighting the proton density regularization terms 4 times as much as the $T_2$ term.

## IV. RESULTS

### A. Test 1: Numerical simulation experiment

For the balanced sequence used in Test 1 and 2, the NRMSEs (Normalized Root Mean Square Error) obtained for the signals and derivatives of the surrogate model are 0.913% and 1.765%, respectively.

For the digital brain phantom reconstructions, we compare the NRMSE of the reconstructed quantitative maps and the cost function values using different ADMM parameters, and here we choose to present the reconstruction results using $\lambda = 2.0$, which shows the fastest convergence. In Fig. 3, the objective function value versus the runtime for 15 iterations are shown for both the proposed and the sparse Hessian reconstructions. It can be seen that both the proposed and the sparse Hessian methods converge to the noise level within 10 iterations. Note that the proposed method will not reach the exact same objective function value as the existing approach, because a surrogate model is used when optimizing the objective function. However, the difference in the objective function values after 15 iterations is relatively small comparing to the initial objective. In particular, the sparse Hessian method reaches a final cost function value of 5% the initial cost, while the proposed method reaches 6%. The difference is therefore negligible. The proposed method takes about 234 seconds for 10 iterations, thereby outperforming the sparse Hessian method by approximately two orders of magnitude. Training of the surrogate model including the generation of the training data (Bloch simulations) takes about 4 hours; however, this step only needs to be executed once for a fixed sequence, and can be re-used for different reconstructions. The $T_1$, $T_2$ and proton density maps ($abs(\rho)$) reconstructed using both methods (10 iterations) are shown in Fig. 4(a), and the mean value of the relative standard deviation (error bars) for three different tissue types are shown in Fig. 4(b). It can be seen in Fig. 4(a) and 4(b) that the parameter maps reconstructed by both methods show similar accuracy. Similar correspondence between the two reconstruction methods was obtained for lower SNR values (results not shown).

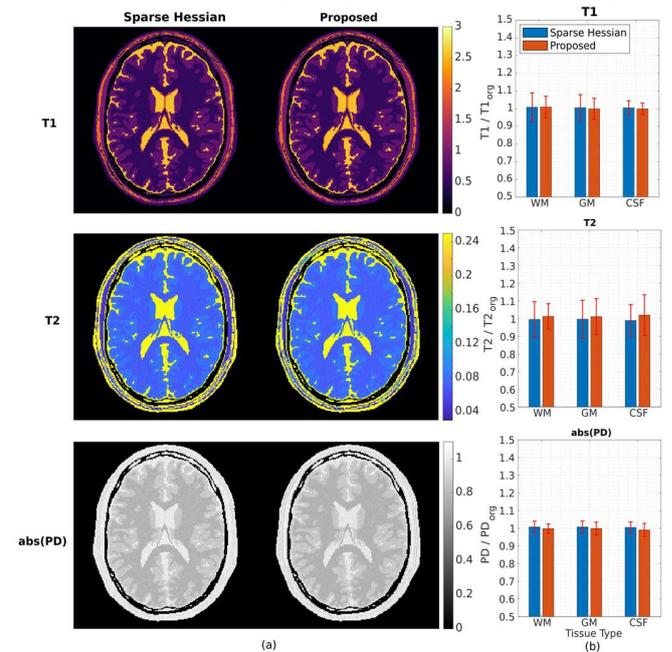

Fig. 4. Reconstruction results of the numerical simulation experiment with SNR = 50. The proposed reconstruction results were compared to the sparse Hessian results. (a) Reconstructed parameter maps after 10 iterations. (b) The mean value and standard deviation (error bars) for three different tissue types, white matter (WM), gray matter (GM) and cerebrospinal fluid (CSF) are shown. The values are normalized by the true parameter values used in the simulation.

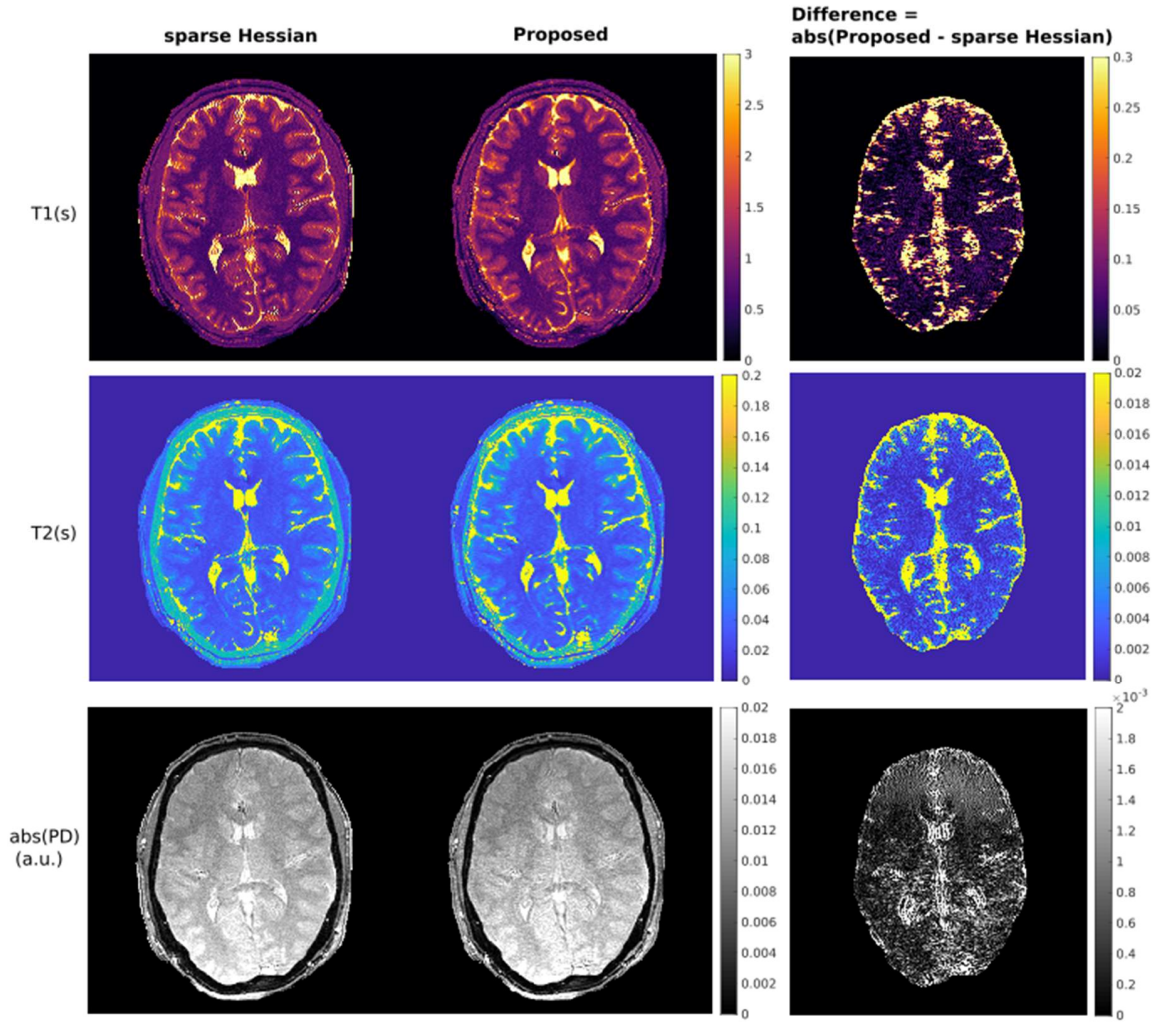

Fig. 5. The $T_1$, $T_2$ and proton density maps, as well as the difference maps, reconstructed from the balanced in-vivo brain data using both the proposed and the sparse Hessian methods are shown. The difference images are plotted in a 10 times smaller range, which reveals negligible difference in white gray matter reconstructed values.

### B. Test 2: In-vivo brain data with the balanced sequence

Fig. 5 shows in-vivo $T_1$, $T_2$ and proton density maps reconstructed by both the proposed and the sparse Hessian MR-STAT reconstruction algorithms. We choose the iteration number for sparse Hessian method to be 6 and for the proposed method to be 7 such that the objective functions converge to similar values. The absolute difference maps are also displayed with a 10 times slower range. It can be seen that the ADMM and sparse Hessian methods return similar solution, and the maximum difference can be observed in the CSF region in the $T_2$ map. CSF values are expected to diverge as a consequence of instability due to much higher relaxation rates and flow effects. The sparse Hessian method took 6886 seconds for 6 iterations, whereas the proposed method took just 171 seconds, achieving an approximate acceleration rate of 40.

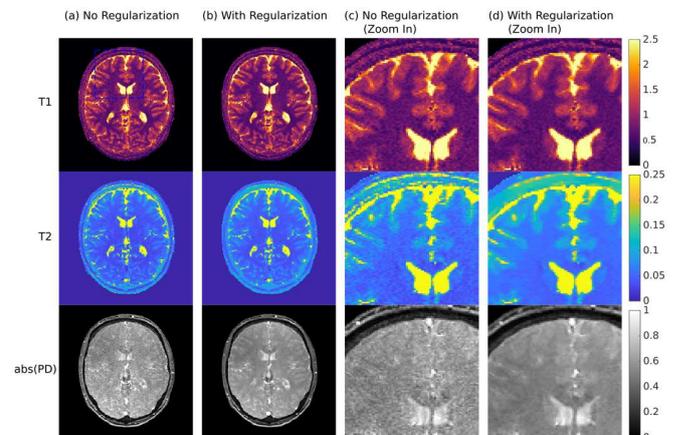

Fig. 6. The T1, T2 and proton density maps reconstructed from the spoiled-gradient in-vivo brain data by the proposed algorithm without (a, c) and with (b, d) the TV regularization terms are shown.

### C. Test 3: In-vivo brain data with the gradient-spoiled sequence

Finally, we applied the ADMM algorithm to reconstruct MR-STAT data acquired with a gradient-spoiled sequence and 4 virtual coils. In Fig. 6, in-vivo $T_1$, $T_2$ and proton density maps reconstructed by the ADMM algorithm with and without the TV regularization terms are shown. For both reconstructions, results after 6 iterations are shown. It can be seen that all parameter maps reconstructed with TV regularizations show better contrast and lower noise level comparing to maps without any regularization, even though the regularization terms are only added to $T_2$ and proton density, and not to $T_1$. Both reconstructions (i.e. with and without regularization) take about 2.5 minutes.

## V. DISCUSSION

In this work, we have presented an accelerated reconstruction algorithm for Cartesian MR-STAT acquisitions. In the proposed algorithm, a neural network is applied as a fast surrogate model for computing the MR signal and its derivatives, and the reconstruction is split into smaller sub-steps by ADMM. In a previous work, high-resolution in-vivo MR-STAT reconstructions were run on a high performance computing cluster using 96 CPU cores [7], and reconstruction of one 2D slice took about sixteen minutes using the sparse Hessian algorithm. In this work, a reconstruction problem of the same size was run on a 8-core desktop PC and took at most three minutes, reducing both the runtime and the computational resources required. Note that the reconstruction time depends on the size of the image mask applied, thus reconstructing upper slices of the brain takes less time than the slices showed in the Results section. Using the proposed algorithm, whole-brain reconstructions of $T_1$, $T_2$ and $\rho$ maps with 30 slices can be reconstructed within one hour on a desktop PC [28], thereby drastically facilitating the application of MR-STAT in the clinical work-flow. Further acceleration of the algorithm could be possibly achieved by parallelization of the algorithm, especially the separable, independent sub-step nonlinear problems, on graphics processing units (GPU). This possibility will be explored in future work.

We compared reconstruction results from our proposed algorithm with the previously developed sparse Hessian algorithm[7]. For the numerical simulation experiment, it can be seen that both algorithms give very similar reconstructions. For the in-vivo experiment with the balanced acquisition, the difference between the two reconstructions are larger than the numerical experiments, especially in the CSF regions. In Figure 5, approximately 10% difference in the bright regions (mostly CSF) can be observed between the proposed method and the sparse Hessian method. In general, in-vivo reconstructions are susceptible to modelling errors, that is, errors caused when using an imperfect signal model. Physics models applied in both reconstruction algorithms neglect the through-plane flow effects of CSF [20], and the error caused by the imperfect physics models may propagate differently in the two nonlinear reconstruction algorithms, leading to the approximately 10% discrepancy in the CSF regions. This 10% discrepancy will not decrease significantly even if more iterations are run. However, the overall brain regions other than the CSF region show every close (approximately 5% or less overall) quantitative values using the two reconstruction algorithms. For the in-vivo experiment with gradient-spoiled sequence, synthetic contrast maps are also generated from the proposed MR-STAT reconstructions (see **Appendix B, Fig. A1)**, showing good agreement compared to the conventionally acquired contrast images. With the proposed algorithm, regularization terms can be easily added, and the corresponding reconstruction provides parameter maps with better SNR at the cost of very little additional runtime. In this work, separate TV regularization terms are added for $T_2$, $re(\rho)$ and $im(\rho)$ maps respectively, neglecting the intrinsic correlations across different parameter maps. Better regularization techniques suitable for multi-parametric reconstructions [29], [30] could be applied to the proposed MR-STAT algorithm, in order to achieve even better image quality.

In this work, we used a neural network model as a fast surrogate for computing the MR signals and derivatives. The small Fully Connected -NN architecture employed in this work has a very simple architecture and is extremely computationally efficient. It can be easily trained with relatively small amount of data, and can compute large amount of MR signals/derivatives in a very short time (less than 5 seconds for a 224x224 image by a CPU). However, one trained NN model could only work for one fixed MR-STAT sequence, and re-training will be required when a new sequence with a new flip-angle train is used. There is always a trade-off between the computation time and generalization capability when choosing surrogate models. We also explored surrogate models with much better generalization capability which could work for sequences with arbitrary RF flip-angle trains, however, the proposed RNN model in [31] is at least one order of magnitude slower than the model we used here. In the future, surrogate models can be trained to more complicated MR physics effects, for example water-fat effects [32] and magnetization transfer [33]. Surrogate models with good generalization capability could also be incorporated for working with different sequence parameters without the need for retraining [31].

In this work, we trained small neural network models with high accuracy for accelerating the MR-STAT reconstructions. Both the accuracy of the predicted derivatives and the magnetization signals from the surrogate model could affect the reconstruction accuracy. However, the accuracy of the surrogate derivatives affects the reconstruction accuracy not as much as the signal accuracy does. In well-established nonlinear optimization problems such as Gauss-Newton, the approximated second-order signal derivatives are used for approximating the Hessian matrix. It is known that many derivative approximation methods, such as finite difference, have been already used in nonlinear optimization algorithms, showing that estimated derivatives can lead to fast and accurate convergence. Based on this experience, we expect that accurate derivative approximations from neural network surrogate models would performs similarly as other approximation method. The reconstruction results in this manuscript represent a practical validation of this strategy. A theoretical proof on how the accuracy of the surrogate derivatives affect the convergence rate of the MR-STAT problem would be beyond the scope of this manuscript.

Recent work, such as the low-rank ADMM (LR-ADMM) MRF algorithm [17], has also incorporated the low-rank property of the MR signal in temporal domain and the ADMM approach for problem splitting. However, our proposed algorithm is conceptually different from the LR-ADMM MRF algorithm, at least in the following aspects. Firstly, our proposed algorithm works for data acquired with Cartesian sampling patterns, whereas most MRF algorithms works for radial or spiral acquisition data. Generalization of our proposed framework to non-cartesian acquisition still requires further investigation [22]. Secondly, our proposed method models the MR signal during the readout period using the physical model, rather than using the Fourier Transform. Therefore, signal evolution during the readout, including $T_2$ decay and off-resonance effect can also be included in our proposed algorithm. In addition, our proposed method splits the problem in a different way compared to the LR-ADMM MRF algorithm. In LR-ADMM MRF, the temporal SVD compression was applied and the nonlinear sub-problem runs the dictionary matching process voxel by voxel using the low-rank signals. In our proposed method, we acquire low-rank signals including readout-encoding effects for each phase-encoding line in the linear sub-step (16), and reconstruct the quantitative maps line by line, not voxel by voxel, in the nonlinear sub-step (17). Finally, the proposed algorithm solves the nonlinear sub-problem by optimization algorithms such as Gauss-Newton algorithm, therefore is a dictionary-free method which yields continuous-valued parameter outputs and it can more easily be applied in case of multi-parameter reconstructions beyond $T_1$, $T_2$ and $\rho$. Since the proposed algorithm is capable of separating the comprehensive MR-STAT problem into smaller sub-steps and has relatively short runtime, it shows great potential to be extended to 3D reconstruction problems. It may also be possible to include more reconstruction parameters, such as diffusion, in the fast surrogate model at the cost of a small increase in reconstruction time, provided that the sequence has sufficient sensitivity for the wanted effects.

## VI. CONCLUSION

A new reconstruction framework incorporating two acceleration strategies is developed for Cartesian MR-STAT acquisitions. A fast surrogate model is used for full and low-rank MR signal and derivatives computation. An ADMM approach is applied to split the problem into linear and nonlinear sub-problems. The resulting algorithm accelerates the MR-STAT reconstruction by a factor of at least 40, and enables reconstructions of one slice at 1 mm × 1 mm in-plane resolution with a desktop PC within three minutes; thereby, the application of MR-STAT in a clinical workflow can be facilitated.

## APPENDIX

### A. Coil sensitivity encoding

Given coil sensitivity maps $g^{[j]}(r)$, the signal model (12) can be slightly modified as

$$S^{[j]}(\alpha) = \int_V g^{[j]}(r) \cdot \rho(r) M(\alpha(r), \beta) dr =$$
$$= \sum_{b=1}^{N_y} C^p(y_b) \cdot U \cdot Y^E(\alpha_b^R) \cdot C^{R,[j]}(\alpha_b^R),$$
(A.1)

where $C^{R,[j]}(\alpha_b^R) = \begin{bmatrix} g^{[j]}_{1,b} \cdot \rho_{1,b} \cdot c^r(\alpha_{1,b}) \\ g^{[j]}_{2,b} \cdot \rho_{1,b} \cdot c^r(\alpha_{2,b}) \\ \vdots \\ g^{[j]}_{N_x,b} \cdot \rho_{N_x,b} \cdot c^r(\alpha_{N_x,b}) \end{bmatrix}$ is the spatially

weighted readout matrix for the $j$-th coil, and $S^{[j]}(\alpha)$ is the measured temporal signal from the $j$-th coil.

Problem (13) can be modified to use multiple (virtual) coil data simultaneously, and a similar ADMM approach as (16-18) is followed for solving the multi-coil problem.

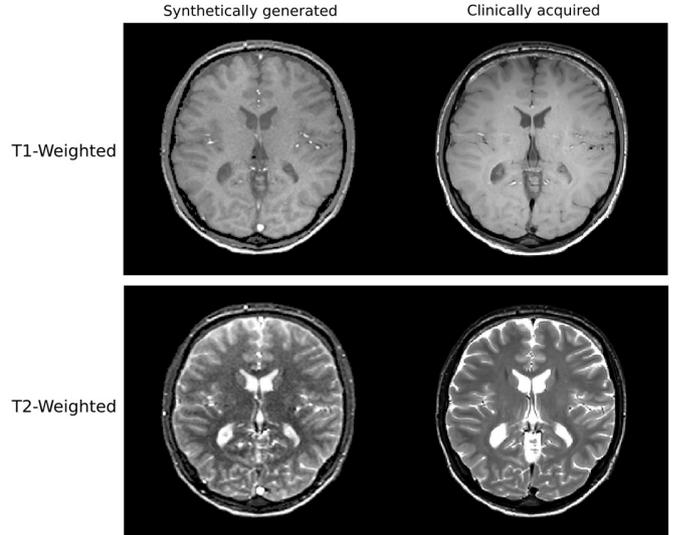

Fig. A1. Comparison between synthetically generated contrast maps (left column) and conventionally acquired contrast maps (right column). Synthetic contrast maps are generated from ADMM reconstructed parameter maps as shown in Fig. 6(b).

### B. Synthetically generated contrast maps from the proposed MR-STAT reconstructions

To further show the clinical potential of the proposed MR-STAT reconstruction, we also generated synthetic contrast images, specifically $T_1$, and $T_2$-weighted images from the reconstructed [28]. We generate the synthetic MR images using the MR-STAT reconstructed images from the gradient-spoiled data, as an extension of Test 3 in the main text. The synthetic MR contrast images were therefore compared with $T_1$, and $T_2$-weighted images acquired by conventional contrast-weighted neuro MRI sequences. The same resolution and voxel size as the MR-STAT experiments were used for the conventional MRI sequences to allow direct comparison. Specifically, conventional $T_1$-weighted image was acquired by a spin-echo sequence with constant TR/TE = 451/14ms and flip angle = 70 degree; conventional $T_2$-weighted image was acquired by a turbo spin-echo sequence with TR/TE = 3400/80 ms, flip angle = 90 degree and TSE = 15.

Synthetic contrast images generated from the ADMM MR-STAT reconstructions with regularization are shown in Figure S1 alongside the conventional T1 and T2 weighted images for comparison. The synthetic MR-STAT T1 and T2 weighted images show good agreement with the conventional contrast images, showing slightly lower SNR but better gray/white matter contrast.

## ACKNOWLEDGMENT

The first author (HL) receives a scholarship granted by the China Scholarship Council (CSC, #201807720088).